\documentclass{article}

\usepackage{lineno,hyperref}
\usepackage[cp1251]{inputenc}
\usepackage{amsfonts,amssymb}
\usepackage{amsmath,amsthm, graphicx, epsfig}
\usepackage[T2A]{fontenc}
\newtheorem{definition}{Definition}
\newtheorem{theorem}{Theorem}
\def\maxl{\max\limits}
\def\minl{\min\limits}
\def\suml{\sum\limits}

\def\supl{\sup\limits}

\newcommand{\df}{\stackrel{\textrm{def}}{=}}
\begin{document}

%\begin{frontmatter}

\title{Binary classification of multi-channel EEG records  based on the  $\epsilon$-complexity of continuous vector functions}

\author{Boris Darkhovsky, Alexandra Piryatinska, Alexander Kaplan}
%\address{Institute for Systems Analysis, FRC CSC RAS,  Higher School of Economics, Moscow, Russia}

%\author{Alexandra Piryatinska\corref{mycorrespondingauthor}}
%\ead{alpiryat@sfsu.edu}

%\author{Alexander Kaplan}
%\ead{www.elsevier.com}

\maketitle

%\address[myaddress1]{Department of Mathematics, San Francisco State University, 1600 Holloway Ave, San Francisco, CA, 94070}
%\address[kaplanaddress1]{Department of Human Physiology
%M.V. Lomonosov Moscow State University, Pirogov Russian National Research Medical University, Moscow, Russia}

\begin{abstract}
A methodology for binary classification of EEG records which correspond to different mental states is proposed. This model-free methodology is based on our theory of the $\epsilon$-complexity of continuous functions which is extended here (see Appendix) to the case of vector functions. This extension permits us to
handle multichannel EEG recordings. The essence of the methodology is to
use the $\epsilon$-complexity coefficients as features to classify (using well known classifiers) different types of vector functions representing EEG-records corresponding to different types of
mental states. We apply our methodology to the problem of classification of multichannel
EEG-records related to a group of healthy adolescents and a group
of adolescents with schizophrenia. We found that   our methodology permits
accurate classification of the data  in the four-dimensional feather space of the $\epsilon$-complexity coefficients.
%{\bf Background and Objective}: The important step in classification of  EEG records is a feature selection. The feature selection problem is difficult due to complex structure of    EEG signals.
%In the modern literature the feature spaces which classify EEG signals with
%good accuracy are high dimensional. Our objective is to create a small dimensional feature space which enables binary classification of EEG records.
%
%{\bf Methods:}
%The proposed approach is based on our theory of the $\epsilon$-complexity of continuous functions which is extended here (see Appendix) to the case of vector-functions. This
%extension permits us to handle multichannel EEG records. The methodology contains two steps. Firstly we  estimate  the $\epsilon$-complexity
%coefficients of original signal and its finite differences. Secondly we  utilize  Random Forest (RF) or Support Vector Machine (SVM) classifiers.
%
%{\bf  Results: }
% We apply our methodology to the problem of classification of  multichannel EEG-records related to a group of healthy adolescents (39 subjects) and a group of adolescents
%with schizophrenia (45 subjects).  We found
%that Random Forest classier provides a superior result. In particular, out-of-bag accuracy in case of RF is 83.6\%. In case of 10 folds cross-validation RF
%gives accuracy 86.2\% on a test set whereas SVM gives accuracy 79.7\%.
%
%{\bf Conclusions:} We developed model-free methodology for binary classification of EEG records. The  feature space was reduced  to four dimension. The obtained results indicate the effectiveness of the proposed methodology.
%
%
%

\end{abstract}

%\begin{keyword}
%\texttt{EEG-analysis}\sep   \texttt{binary classification} \sep \texttt{$\epsilon$-complexity} \sep
%\texttt{schizophrenia}
%
%\end{keyword}

%\end{frontmatter}

%\linenumbers

\section*{Introduction}

An electroencephalogram (EEG) is a direct measure of  electrical activities of the brain along the scalp.
  It  is a rich source of information about the brain for  healthy  individuals and patients with neurological diseases.
  Compared  to the blood flow neuroimaging techniques, such as magnetic resonance imaging (MRI) and functional magnetic resonance imaging (fMRI), which are indirect measures of brain activity,    EEG  is  cheaper and easier to use.  An additional strength   of EEG   is that it can detect changes within a millisecond time frame whereas fMRI has time resolution between seconds and minutes.
 Quantitative evaluation  of cognitive functioning and mental states using EEG
records is one of the important  problems in  applied psychophysiology \cite{haynes2006decoding}. They include   brain-computer interface \cite{aflalo2015decoding, kaplan2005unconscious, Kaplan2013} for decoding   intentions and their translation  into commands, and diagnosis  of mental illnesses such as  schizophrenia \cite{haynes2006decoding, loo2015research, neto2015eeg}.

 To obtain useful information from the EEG data feature extraction is necessary. In the literature  a whole set of quantitative estimates of the  spectral and temporal features
 of the EEG signal  have been  developed (see, e.g. \cite{van2007application}).
 In particular, there is an extensive literature on attempts to use these  characteristics  for schizophrenia diagnosis (for review and meta-analysis of such papers see e.g. \cite{boutros2008status}). However such approach requires  assumptions on a data generating mechanism but there are no generally excepted model on the data generated mechanism for EEG data.

  Another approach to select features is related to
the fundamental concepts of the modern theory of nonlinear dynamical systems
such as entropy, correlation dimension $D_2$ and Lyapunov exponent (see, e.g. \cite{babloyantz1986evidence, roschke1992strange, pereda, kaplan2001}). These features  can really reflect the complexity of a generating mechanism
   of a signal, but only under the   assumptions of    stationarity and ergodicity of a signal.   Due  to the nature of EEG signals these assumptions   are not fully justifiable (see, e.g. \cite{subha2010eeg, Kaplan1998Nonstationarity}).

To the best of our knowledge  the papers which produce a high accuracy on classification of EEG signal use high-dimensional feature space  and face the problem of overfitting (see e.g. \cite{dvey2015schizophrenia})

In this paper we propose an approach to the feature selection problem which
 is model free (e.g. does not have any assumptions on data generating mechanism  such as stationarity or ergodicity of the data).  It also  assures a good accuracy  in small dimensional feature space.	More precisely, we propose to apply the notion of the $\epsilon$-complexity of continuous functions  to the classification problem of
 multichannel EEG-records.
	Such approach is in line with general  Kolmogorov's idea on a ``complexity'' of an object. At the basic level the idea of a  Kolmogorov can be expressed as follows: A ``complex'' object requires a lot of information for its reconstruction and, for a ``simple'' object, little information is needed. Therefore, it is quite natural to measure
the complexity of a continuous function by the number of function values on a uniform grid (e.g. function values given at equally distant points) which are necessary to reconstruct the function with a given error by a given set of methods.
	This characteristic of complexity was tested on evaluation of the functional states of the brain using the EEG recordings in  \cite{dark2002,darkhovsky2006estimation}. It was found that the mean values of complexity are statistically significantly different for the groups of subjects with different functional states. This result was obtained despite the fact that the easiest method of function reconstruction by piecewise-constant approximation was used. However, for further progress in this area the development of computational  procedure for estimation of  the ``complexity'' of an \emph{individual recording} (i.e.  an  \emph{individual continuous function}) was required. This was  impossible without development of an appropriate mathematical theory.

In 2012-2014 ( see, \cite{darkhovsky2012new, darkhovsky2014new, darkhovsky2014quickest,  darkhovsky2015novel}) the theory of the $\epsilon$-complexity of continuous functions defined on a compact set in a  finite-dimensional space was developed.  It  was proven   that the $\epsilon$-complexity of  ``almost all''  H\"{o}lder functions is effectively characterized by a pair of real numbers, which we call the \emph{$\epsilon$-complexity coefficients}.
This  theory enabled us to   develop  a \emph{novel  approach} to the problems of segmentation and classification of time series of arbitrary nature.
In this paper  we extended    the theory  of the $\epsilon$-complexity of continuous functions   to the  case of continuous vector-functions (see Appendix). This extension enables us to apply such approach to the binary classification problem of EEG records.

The paper is organized as follows. In  Section 1   we describe our methodology. In particular,  in Subsection  1.1 we give  a description of the notion of  the $\epsilon$-complexity of a vector function  on a semantic level and  provide a characterization of the $\epsilon$-complexity  for vector functions given by a finite set of values.   In Subsection 1.2 we provide an algorithm  for estimation of the $\epsilon$-complexity coefficients  for the multichannel EEG records. In Subsection 1.3 we describe our classification procedure.  In Section 2 we apply our methodology  to the  classification of the EEG records of adolescents   with schizophrenic type of disorder and of healthy subjects.  We have established that  these data permit accurate classification in the four-dimensional space of the $\epsilon$-complexity
coefficients of original EEG signals and $\epsilon$-complexity coefficients of fourth differences of  EEG signals. In Section 3  we provide conclusions and discuss  our results. The Appendix provides the precise definition of $\epsilon$-complexity and the theorem characterizing the complexity of H\"older vector functions

\section{Methodology }

In this section, we give  a    description  of proposed methodology    for  classification of multi-channel EEG-records.

We will  treat a  multichannel EEG record as  a
 $d$-dimensional vector function  $x(t)=\left(x_1(t),\dots, x_d(t)\right)$, where  $d$ is a number of channels,
  which is given on some  fixed time interval  $t\in [0,T]$ .

Since the modern recording equipment is digital, instead of a continuous vector function  $x(t)$ the researcher
 received a discrete samples $\left(x(0),x(T/n),x(2T/n),\dots,x(T)\right)$, e.g. , the sequence  of  $n$ $d$-dimensional vectors.  Here $n=fT$, where $f$ is a sample frequency (if  the frequency is measured in Hz,  e.g. time in seconds). For example, if  $T=60$ seconds, and  $f=128$Hz, then we have  7680 $d$-dimensional vectors.

Without loss of generality we can assume that
 $\maxl_{0\le k\le n}|x_i(kT/n)|=R_i>0,\,i=1,2,\dots,d$, e.g. in each channel of EEG signal is present.

\subsection{Description of the $\epsilon$-complexity of continuous vector function}

Let us describe our notion of the $\epsilon$-complexity on semantic level. The precise definition and formulation of the theorem are given in Appendix.

We  choose a number  $0<S<1$  and discard from each component of a vector function     $\{x_i(kT/n)\}_{k=0}^{k=n},\,i=1,\dots,d$   $[(1-S) n]$ values  (henceforth, the symbol $[a]$ denotes the integral part of a number $a$), in such a way that
the remaining values are approximately uniformly distributed. For example, if $S=0.5$, then we retain even or odds  values  in each component of the function.

Assume we have some fixed collection $\mathcal{F}$ of approximation methods which can be used for reconstruction of a continuous function by its values at some uniform grid. Employing   collection of methods  $\mathcal{F}$, we reconstruct the values of each component of the vector function   $i,\,i=1,\dots,d$ in discarded points using the retained values of the  function. We find the methods which reconstruct the function with minimum relative (in relation to  $R_i, i=1,\dots,d$) error  (the error
can be measured in any norm, because we are dealing with a finite a set of values).

Denote  the value of the minimum relative reconstruction error in the $i$-th component through
 $\epsilon_i(S)$ and find the value $\epsilon(S)=\suml_{i=1}^d \epsilon_i(S)$.

Now we will \emph{define the  $(\epsilon,\mathcal{F})$-complexity (hereafter, for simplicity of presentation we will write
 $\epsilon$-complexity)  of continuous vector function
  $x(t)$, which is given by its values on the uniform grid by $(-\log S)$}.

 In other words, the  $\epsilon$-complexity of a vector function is the (minus) logarithm of relative fraction of their values, required for its recovery by methods  from family $\mathcal F$ with a relative error no more  than $\epsilon$. In other words, it is ``the shortest'' description of the vector function.
 %Therefore our definition  is in line with the Kolmogorov's idea mentioned in the introduction.

Let us consider the class of vector functions satisfying the H\"{o}lder condition.
 It means that for any  $(t,s)\in [0,T]\times [0,T]$  the following inequality holds
\begin{equation}\label{eq1}
\suml_{i=1}^d |x_i(t)-x_i(s)|\le L|t-s|^p,\,\,L>0, p>0.
\end{equation}

This class of vector functions is very wide, and it includes
practically  all vector functions which can be found  in applications.

The main point of our classification methodology is as follows. For
``almost all'' vector functions satisfying the H\"{o}lder condition, in case of  sufficiently rich family $\mathcal{F}$ of  approximation  methods and a sufficiently large  sample size $n$ there exists range $0<\alpha(n)\le S\le\beta(n)<1$ (which depends from vector function) such that the following equality holds
\begin{equation}\label{eq2}
\log \epsilon\approx A + B\log S
\end{equation}

The precise  meaning of the expression ``almost any'' and symbol
$\approx$ will be explained in the Appendix.

The above parameters  $A, B$ are called the  \emph{$\epsilon$-complexity coefficients}.
These  $\epsilon$-complexity coefficients will be  utilized as features for   classification of multichannel EEG records. In our work  we will use supervised classifiers such as random forest and support vector machine.

These features don't depend from the data generating mechanism and therefore they are \emph{model-free}. In the
scalar case they have been introduced and applied for the purpose to detect changes in
generating mechanism \cite{darkhovsky2012new, darkhovsky2014new, darkhovsky2014quickest,  darkhovsky2015novel}.

We would like to mention two facts. Firstly, in practical applications the family of
function reconstruction methods $\mathcal{F}$ is finite.
It follows from our main   theorem (see Appendix) that  if this family $\mathcal{F}$ is reach
enough and the sample size of function values is large enough then the error of
vector function reconstruction in discarded points by methods from this family (and therefore
the $\epsilon$-complexity) is closed to the error of the reconstruction by \emph{all (computable )
methods}.

Our experience with simulations and real data shows that the dependence
(\ref{eq2}) has been observed in case we chose piece-wise polynomial functions up to fourth
degree as family $\mathcal{F}$.

Secondly, it is not significant which method gives the smallest
error of reconstruction $\epsilon_i(S)$ in $i$-th channel ($i=1,\dots, d$). To find dependence (\ref{eq2}) we need only values of minimal errors.

\subsection{Algorithm for estimation of the $\epsilon$-complexity coefficients}
In this subsection we will  describe main steps of our algorithm  for estimation of the $\epsilon$-complexity coefficients for  multichannel  EEG records.

\begin{enumerate}
 \item
 Normalize each component of the EEG record  $x_i(t)$, i.e
 replace our original  components of the multichannel record  by $x_i(t)/\max_t(|x_i(t)|)$.
 \item
 Select $S$, the fraction   of the remaining points
 as follows: $S_1=50\%$, $S_2=33\%$,  $S_3=29\%$, $S_4= 25\%$, $S_5=22.5\%$, $S_6=20\%$.
  \item
 For each fixed $S_i$ ($i=1,\dots, 6$) and for each component of the  multichannel record   discard  the values of  the functions at points which are placed uniformly,  or almost uniformly,  according to the following scheme:
   Let $x_i^1, x_i^2$,$x_i^3,
  \dots, x_i^n$ be the   values of a  function on a  grid.
  \begin{enumerate}
  \item
     {\bf  ${S}_1=50\% $}: Values of
        $x_i^2, x_i^4, \dots, x_i^{2j}, \dots;$  or  $x_i^1, x_i^3, \dots, x_i^{2j+1}, \dots;$
        are discarded. Notice we have two different  ways  to discard function values;
        \item
          {\bf  $S_2=33\%$} :Values of $x_i^1,x_i^4, x_i^7,x_i^{10}, \dots$; or $x_i^2,x_i^5, x_i^8,x_i^{11}, \dots;$ or $x_i^3,x_i^6, x_i^9,x_i^{12}, \dots;$ are discarded. We have three different placements of discarded values;
          \item
         {\bf  $S_4=25\%$} : Values of $x_i^1,x_i^5, x_i^9,x_i^{13},x_i^{17}, \dots;$ or $x_i^2,x_i^6, x_i^{10},x_i^{14},x_i^{18},\dots$;\\ or $x_i^3,x_i^7, x_{11},x_{15},x_{19}, \dots;$ or $x_4,x_8,x_{12}, x_{16},x_{20}, \dots;$  are discarded. We have 4 different placements of discarded values;
                          \item
  The procedures are similar     in the case  $S_3=29\%$,  $S_5=22.5\%$ and $S_6=20\%$.
  \end{enumerate}
\item
For each $S_k$ and for each  of those placements we consider all possible reconstructions of the function by piecewise polynomials up to fourth degree and  select the one which provides the minimal error of reconstruction. Record this value of the minimal error.
   \item
   For the same $S_k$ we consider other placements of the retained points and repeat the procedure. Record the obtained minimal  errors.
   \item
   Then we take a mean of the recorded errors calculated   over all placements for each channel of the EEG-record.
      \item
   Take sum of the mean errors over all channels.
It is  our  estimation of $\epsilon_k$   in  the case of  $S_k$.
\item
Repeat the procedure for  $k=1,\dots,6$.
\item
Consider  points   $(\log(S_k),\log(\epsilon_k))$, and find the best linear fit
\begin{equation}\label{eq3}
 \log\epsilon \approx A + B \log S
\end{equation}
 using the least squares method.
  \end{enumerate}

\begin{figure}
\includegraphics[width=0.45\textwidth]{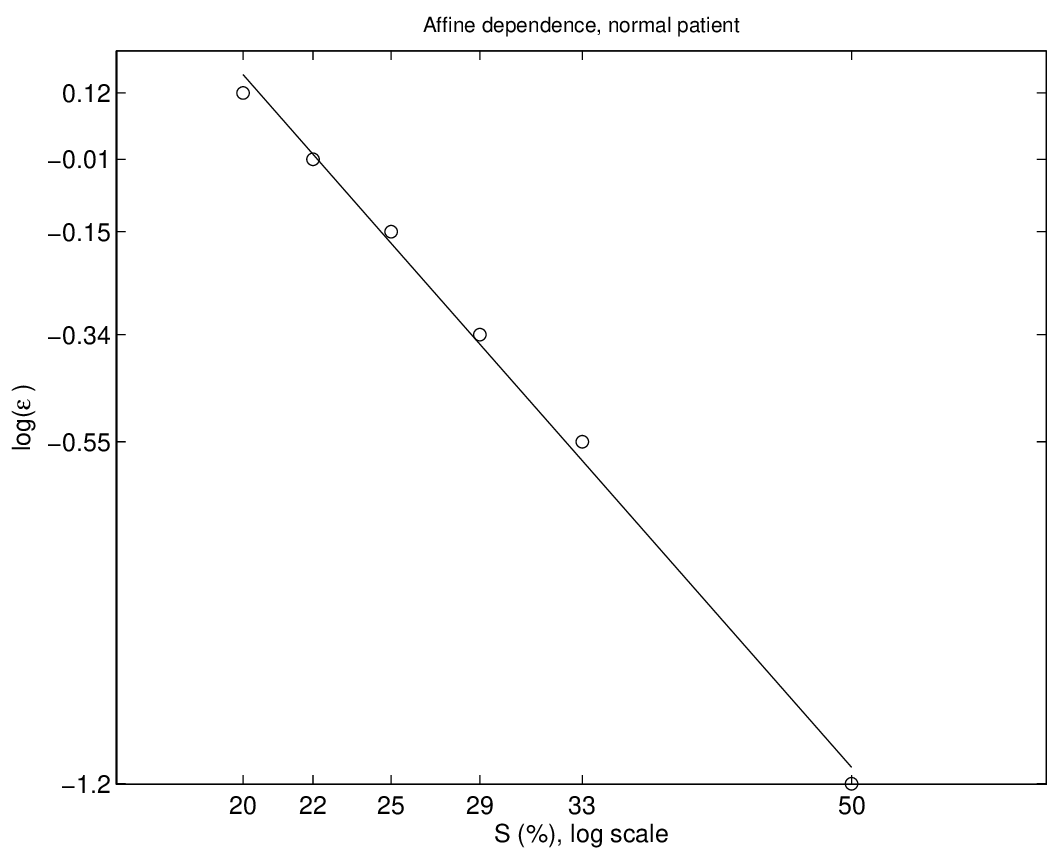}
\includegraphics[width=0.45\textwidth]{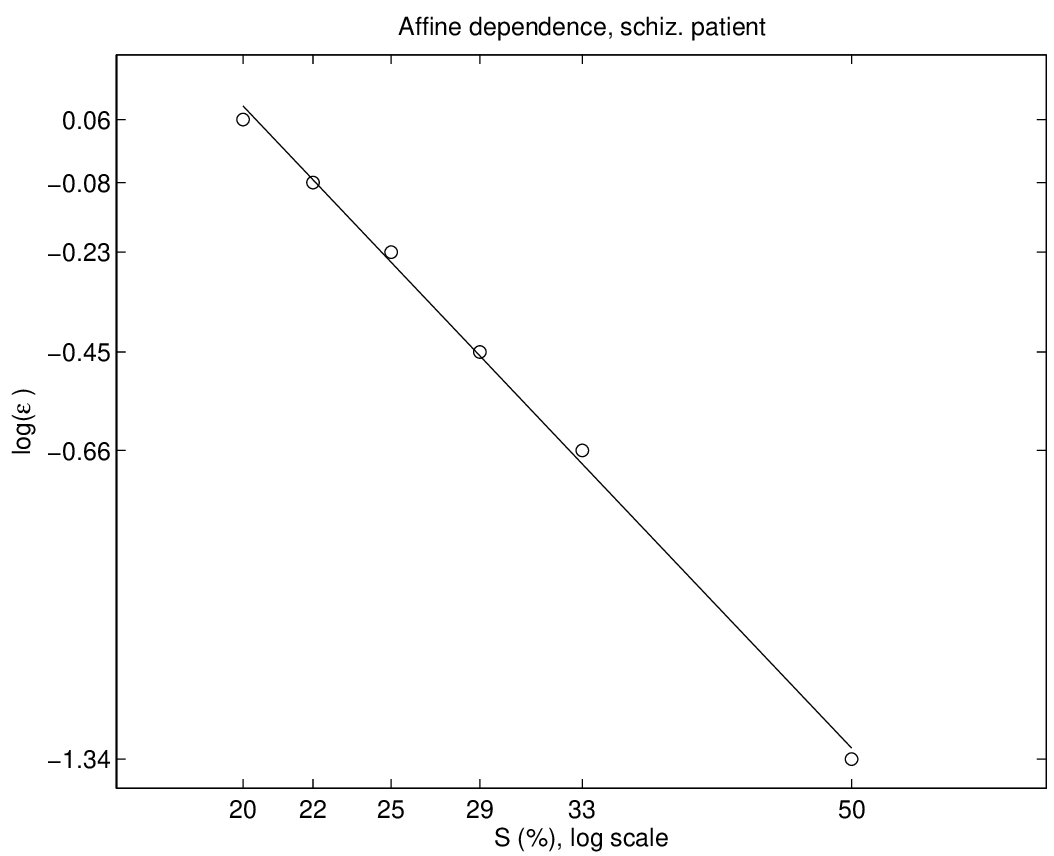}
\vspace{1in}
\caption{Typical dependencies of the form (\ref{eq3}) of EEG  recordings, left (A): normal subject, right (B): patient with  schizophrenia. The theoretical dependence is shown by
a solid line, and the experimental points are shown by circles.
}
\label{line1}
\end{figure}

Each graph  in Fig \ref{line1} demonstrates the typical dependence of the form (2) for given EEG  recording (the theoretical dependence is shown by
a solid line, and the experimental points are shown by circles). The left plot corresponds to the normal subject and the right one to the patient with schizophrenia.

Let us notice that the obtained above values of the coefficients $A$ and $B$ are our estimates for the $\epsilon$-complexity coefficients which are used in classification algorithm.  For each subject we  estimate the $\epsilon$-complexity coefficients as well as the $\epsilon$-complexity coefficients of some transformations of the EEG data (see next section).

\subsection{ Classification }
 In the next step we  use the calculated  $\epsilon$-complexity coefficients as an input to the supervised classifiers, such as Random Forest and Support Vector Machine. The results will be evaluated using the Out-Of-Bag (OOB) \cite{breiman2001random} error in case of the random forest and k-fold cross-validation procedure in both cases to be able to compare the performance of two classifiers.

Note that in case of OOB the model calculates the error using observations not trained  for each decision tree in the forest and aggregates over all of them;  therefore it has no  bias. This is considered to be  an accurate estimate of the test error for the Random Forest \cite{breiman2001random}.

Cross-validation (CV), (see e.g,\cite{geisser1993predictive, kohavi1995study} ) is a model validation
technique for assessing how the results of a statistical analysis will be generalized  to a new data set.
In particular, the data set is  partitioned into complementary subsets (so called  training and validation sets).
The classifier is built using the training data set  and performance is tested on the validation set.
In $k$-fold  cross-validation (see, e.g \cite{kohavi1995study, efron1997improvements}),
the original sample is randomly partitioned into $k$ equal-sized subsamples. Of the $k$
 subsamples, a single subsample is retained as the validation data for testing the classifier,
  and the remaining $k-1$ subsamples are used as training data. The cross-validation process
  is then repeated $k$ times, with each of the $k$ subsamples used exactly once
  as the validation data. The $k$ results from the folds are  averaged to produce a single estimate.

%\mathfrak{}

% Results and Discussion can be combined.

\section{Results}

{\bf Data description.}
EEG recordings  which were analyzed previously  in \cite{borisov2004segmental, ala2005classification,borisov2005analysis} are also used in this study.
The data are publicly available at \url{http://brain.bio.msu.ru/eeg_schizophrenia.htm}. The recordings have been  obtained for 45 boys (10-14 years old) suffering from schizophrenia and diagnosed using clinical interviews at the Research Center for Psychological Disorders of the Russian Academy of Medical Sciences according to the criteria given  in \cite{md1994diagnostic}. Because the patients  had not taken psychoactive drugs before participating in the study, we could exclude the influence of medications.
The control group consists of 39  healthy boys (11-13 years old). The traditional approach to placement of electrodes used at     Pirogov Russian National Research Medical University
in similar studies was employed.
 For all subjects EEG were registered using the standard 10/20 international electrode scheme involving 16 electrodes (O1, O2, P3, P4, Pz, T5, T6, C3, C4, Cz, T3, T4, F3, F4, F7, F8)  with the reference earlobe electrode.

Artifacts, mostly head and eyes movements, were removed manually based on the opinion of two experts. We received data which are free of the artifacts. During the recordings patients were  in resting state with closed eyes. Impedance for all electrodes was kept below 10 k$\Omega$. Signal was sampled at frequency 128Hz and bandpass filter  between 0.5Hz and  45Hz was applied.
The length of each recording  after removal  of the artifacts is   7680 points.

\textbf{Remark. }
The  bandpass-filter between  0.5 and 45Hz was performed initially.
We did not perform any additional  filtration of the signal in different frequency bands. Note the filtration changes the signal and therefore, generally speaking  its $\epsilon$-complexity. However in our attempt to classify the EEG signals into schizophrenic and control groups we did not have a preliminary information
which frequency band can be used as classification feature. Therefore we did not perform such filtration and as it turned out later, we  need not it.

In our analysis the  multichannel EEG record is treated  as a restriction  of a continuous  vector function $x(t)= \left(x_1(t),\dots,x_{16}(t)\right),t\in [a,b]$ at the uniform grid, i.e. values  of continuous vector function are given in equally distant points of time  with distance 1/128 sec.

For all individuals in our study we estimated the $\epsilon$-complexity coefficients $A_i, B_i$ according to the above algorithm. Here, $i$ is the subject's number, $i=1,\dots, 84$, where the first 39 subjects are normal and the last 45 subjects  are schizophrenic.

After that we consider differences of the   original EEG signals,  $x^{(1)}(t)=x(t+1)-x(t)$, $t=1,..,n-1$,
$x^{(2)}(t)=x^{(1)}(t+1)-x^{(1)}(t)$
$t=1,....,n-2$,  and etc.
The $\epsilon$-complexity coefficients $ADk_i, BDk_i$, $k=1,2,3,4$ are estimated up to the  fourth difference.

The employment of differences corresponds to the analysis of derivatives of the  EEG signal. Since,  a priory , we did not know the features of the signal which are useful
for classification,
we tested  different combinations of the $\epsilon$-complexity coefficients of the original series and the  series of finite differences and found that the complexity coefficients  $ A_i, B_i$, and complexity coefficients of 4th  differences $AD4_i, BD4_i$ ($i=1,\dots, 84$) form   the  best  set of features for classification of our patients into two groups: schizophrenic and control.

After that  we use the     machine learning techniques
    to separate data into  two clusters.
    In particular,  we feed the vectors $(A_i, B_i, AD4_i, BD4_i)$ into the supervised  classifiers such as    Random Forest (RF) and
     Support Vector Machine classifiers (SVM).    This step is performed using R project software and       its  package "RandomForest" for the random forest classifier and package "e1071" for the SVM. We  also perform 10-fold cross-validation  using the  "cart" package.

The results for the OOB   and  the $10$-fold cross-validation for
  Random Forest and SVM classifiers are presented   in the Table 1. In this table we provide the accuracy of these classifiers and the
   percentage of  false negative and false positive cases. False positive cases are the cases in which we classify the healthy
  patient as a patient with  schizophrenia and
  false negative cases  are the cases in which  we classify  a patient with schizophrenia as a healthy patient. To get 95\% bootstrap confidence intervals (CI)  we  performed  10,000 replications of our experiments using random re-sampling of the data.
    %  The histogram of percentage of  accuracy, false negative and false positive cases  over 10,000 replications of 10-folds CV  in case of Random Forest  are  presented in Fig. \ref{hist}.

\medskip
Table 1. (Results for the Random Forest (RF) and Support Vector Machine (SVM) classifiers)

\medskip
%\medskip
{\small
\begin{tabular}{|l|l|l|l|l|}
  \hline
   Method& &  Accuracy (in $\%$)& False Positive (in $\%$) & False Negative (in $\%$) \\ \hline
  RF &OOB & 83.6 & 11.6 & 20.7\\
RF&95\%  CI & (80.9, 85.7) & (10.3, 15.4)  & (17.8, 22.2)\\
 RF&    10-fold CV& 83.2\% &13.1\% &21.5\%\\
RF&95\%  CI &  (79.2,86.8) &(7.5, 20.0)      &(16.0,27.0)\\
  \hline
  SVM &    10-fold CV& 79.7\% &13.6\% &25.9\%\\
SVM& 95\% bootstrap CI  &  (75.9,83.4) &(8.3, 19.2)      &(20.5,31.7)\\
  \hline
\end{tabular}
}
\medskip

%
%
%\begin{figure}
%\includegraphics[width=0.25\textwidth]{hist_acrf.eps}
%\includegraphics[width=0.25\textwidth]{hist_fprf.eps}
%\includegraphics[width=0.25\textwidth]{hist_fnrf.eps}
%\vspace{1in}
%\caption{ Histograms in 10-fold cross-validation (CV) with 10000 replications. Left: Accuracy ,
%Middle: False Positive,
%Right: False Negative
%}
%\label{hist}
%\end{figure}

\newpage

\section{Conclusions and Discussion}

Until now the problem of  detecting EEG mental states in humans and in particular, in the diagnosis of mental diseases, remains open \cite{van2007application, dvey2015schizophrenia}.
This is  due to the lack of effectiveness of traditional methods of EEG analysis in relation to the classification of normal and pathological mental states \cite{badrakalimuthu2011eeg, kim2015diagnostic}  that causes the researchers to seek new methods of quantitative evaluation of the EEG including  schizophrenia detection and classification \cite{dvey2015schizophrenia}.

In this paper, we proposed  methodology which employs  the $\epsilon$-complexity for the separation of different mental state.  In particular, we apply this methodology to separate EEG records of patients with schizophrenia and control group.

We selected the multivariate $\epsilon$-complexity coefficients of the original data as well as  the multivariate $\epsilon$-complexity coefficients of the  fourth differences as markers in  identification of   schizophrenia in adolescents.   We found that Random Forest classifier  performs
 better for this data set  than Support Vector Machine.
The accuracy of proposed markers   is relatively high, reaching on average $83.6\%$ in the test set used in the cross-validation  on a sample of 45 schizophrenic patients and $39$ healthy subjects. The False Positive  error rate on average is 11.6\% and False Negative is 20.7\%.

We would like to emphasize that the feature space used in classification algorithm has \emph{only four dimensions} which is much smaller than the sample size.
To the best of our knowledge,
 the literature does not contain any results on the classification of
schizophrenic diseases in a space of features of EEG data of such small dimension.

 In the previous work on these EEG data \cite{ala2005classification} six spectral and four time domain characteristics for 16  channels were considered as features for classification. In total 160 features were tested.  It was found that 38 characteristics are sufficient for the separation of two groups of  subjects: schizophrenic and normal.

In this paper we analyze EEG data for  patients  in resting condition without  medical treatment.
There are other known approaches to diagnose    schizophrenia  using EEG data(\cite{dvey2015schizophrenia,light2015validation}), which are based on using different stimuli and post-treatment EEG in different groups of patients. This is a fundamentally different approach, and it is difficult to produce  a comparative analysis of our approach with these approaches. The proposed  methodology expands the arsenal of methods for  classification of EEG signal  into groups of  schizophrenic and normal subjects.

Let us emphasize that  this paper is the  first attempt to  apply the $\epsilon$-complexity of vector functions to real EEG data. Preliminarily, we verified this methodology on simulations but such  results are not presented in this paper. In the future we will apply such methodology to other types of studies and classification problem for EEG signals, such as identification  of pre-seizer states for epileptic patients.

Of interest is  further development of the proposed  methodology for the detection of fixed mental states (Motor Imagery) in real time in the context  of brain-computer interface \cite{hohne2014motor}.
%This methodology can be also useful as supplementary to known methodologies such as interviews of the patients.

\section*{Acknowledgments}
This study was partially supported by funding from the Skolkovo Foundation (project $\sharp$ 1110034) and from Pirogov Russian National Research Medical University. We would like to thank Dr. Gorbachevskaya N.L. (Senior Researcher as Russian Mental Health Research Center, Moscow),   Dr. Borisov S.V. (Senior Researcher at Faculty of Biology M.V.Lomonosov Moscow State University) for collection  and preliminary  processing of the data.

\section*{Appendix}

Let us provide precise definitions and results following from the general $\epsilon$-complexity theory of continuous  functions (see, \cite{darkhovsky2014new}).

Consider  a continuous vector function, $x(t)=\left(x_1(t),\dots, x_d(t)\right)$,   defined on $[0,1]$. Let $R_i\df \maxl_{t\in [0,1]}|x_i(t)|,\,i\in I\df\{1,\dots,d\}$.
We will assume that $$\minl_{i\in I} R_i>0$$.

%(we will use the following norm for an arbitrary function on $[0,1]$:
%$\|x(\cdot)\|_{\infty}\df\suml_{i=1}^d\supl_{t\in [0,1]}|x_i(t)|$; evidently that it is possible to use any other norm in finite-dimensional space).

Let $\hat{x}_i(\cdot)$ be an approximation of the $i$-th component $x_i(\cdot),\,i\in I$ of the vector function $x(\cdot)$ constructed using  its values at the nodes of a uniform grid with spacing $h$ by  one of the allowable methods  of function reconstruction from  a  given  collection   of approximation methods $\mathcal{F} $ .

The function $x_i(\cdot)$ is called \emph{$\mathcal{F}$-nontrivial (correspondingly, totally nontrivial)} if it can not be recovered with an arbitrary small error by methods  $\mathcal{F}$ (correspondingly, by any enumerable collection of methods) for any $h>0$.

The vector function $x(\cdot)$ is called \emph{$\mathcal{F}$-nontrivial (respectively, totally nontrivial)} if all its components are \emph{$\mathcal{F}$-nontrivial (respectively, totally nontrivial)}.

Denote by $\tilde{I}\df\{i\in I: x_i(\cdot) \,\mbox{is an}\,\mathcal{F}-\mbox{nontrivial function}\}$, and
put
$$
\delta_{i}^{\mathcal{F}}(h)=\inf_{\hat{x}_i(\cdot)\in\mathcal{F}}\supl_{t\in [0,1]} |x_i(t)-\hat{x}_i(t)|,\,i\in I.
$$
The function   $\delta_{i}^{\mathcal{F}}(h)$ is called  the \emph{absolute recovery error of the component  $x_i(\cdot)$ by methods $\mathcal{F}$}.
For each $\epsilon \ge 0$, define
%\begin{equation*}
$$
 h^*_x(\epsilon,\mathcal{F})=\left\{\begin{array}{ll}
                        \inf\{h\le 1:\suml_{i\in\tilde{I}}\frac{\delta_i^{\mathcal{F}}(h)}{R_i}>\epsilon \},& \text{if}\,\tilde{I}\neq\emptyset\\
                        1,& \text{in opposite case}
                        \end{array}
                        \right.
 $$
%\end{equation*}
We will call $\frac{\delta_i^{\mathcal{F}}(h)}{R_i}$ \emph{the relative recovery error of the component $x_i(\cdot)$ by methods $\mathcal{F}$}.

\begin{definition}
The number
$$
\mathbb{S}_x(\epsilon, \mathcal{F})=-\log h^*_x(\epsilon,\mathcal{F})
$$
  is called the $(\epsilon, \mathcal{F})$-complexity
  of an individual continuous vector function
   $x(\cdot)$.
\end{definition}

This definition is a generalization of the  main definition for  scalar functions introduced in   \cite{darkhovsky2014new}.

In most modern applications, one deals with  functions (vector functions) defined by their values at a discrete set of equally distant moments of time  (or as it is called, on uniform grid). We will assume that this array of values is the restriction of a continuous function to the points of some uniform grid.

Let $\mathcal{T}$ be a set of totally nontrivial vector functions satisfying   the H\"{o}lder condition, which means that for any $(t,s)\in [0,1]\times [0,1]$

$$
\suml_{i\in I}|x_i(t)-x_i(s)|\le L|t-s|^p,\,L>0,\,p>0.
$$

It can be shown that $\mathcal{T}$ is everywhere dense in the set of  vector functions satisfying  the H\"{o}lder condition. In other words, ``almost any'' H\"{o}lder vector function is totally nontrivial, i.e.  has totally nontrivial component
s.

In our context a  vector function satisfying the H\"older condition is given by its $n$ values (i.e. by $n$ vectors from $\mathbb{R}^d$) on a uniform grid. We choose $0<S<1$ and discard $[(1-S) n]$ of  the function values from    each component of the vector function sample.
Using the remaining  values we  approximate the values of  the  components of  the vector function at the discarded
points by the set of approximation methods $\mathcal{F}$, and for each component find the best approximation in the sense of minimal relative recovery error. Let the minimal relative recovery error of the $i$-th component be  equal to $\epsilon_i,\,i\in I$. Define the relative error of the vector function approximation as $\epsilon=\suml_{i\in I}\epsilon_i$.

Following the main idea, we define now the $\epsilon$-complexity of continuous vector function, given by its values at a  uniform grid,
 as (minus)  logarithm of relative fraction of their values (i.e., $-\log S$),
 which should be retained to reconstruct this function in discarded points with relative error not large then the  $\epsilon$.
 It is easy to show that the definition of
the  $\epsilon$-complexity of function given on discrete set of points converges to the $\epsilon$-complexity of the corresponding continuous function with the increasing
  of sampling frequency   (see \cite{darkhovsky2015novel}).

Due to the fact that a  vector function has a finite number of components and each component  satisfies   the H\"{o}lder condition, from the general theory (see \cite{darkhovsky2014new}) we can immediately obtain the following
\begin{theorem}
For any vector function $x(\cdot)$ from a dense subset of set $\mathcal{T}$, any (sufficiently small) $\kappa>0,\delta>0$, and $n\ge n_{\scriptscriptstyle 0}(x(\cdot))$ there exist  a set of approximation methods $\mathcal{F}^*$, numbers $0<\alpha\left(n,x(\cdot)\right) < \beta\left(n,x(\cdot)\right)<1$, functions $\rho(S),\xi(S)$ and a set $N\subset Q=[\alpha(\cdot),\beta(\cdot)],\mu(N)>\mu(Q)-\delta$ ($\mu(\cdot)$ is Lebesgue measure) such that for all $\mathcal{F}\supseteq\mathcal{F}^*$ and $S\in N$ the following relations hold:
$$
\log\epsilon=A+\left(B+\rho(S)\right)\log S+\xi(S), \,\sup_{S\in N}\max(|\rho(S)|,|\xi(S)|)\le\kappa.
$$
\end{theorem}

It follows from this theorem  that (in the case of sufficiently rich family of approximation methods  $\mathcal{F}$ and sufficiently large $n$) for
vector functions satisfying the H\"{o}llder condition and defined by their $n$ values at a uniform grid the $\epsilon$-complexity is characterized by a pair of real
numbers $(A,B)$ via the formula
\begin{equation}\label{linear}
\log \epsilon\approx A + B\log S.
\end{equation}

Here the notation $\approx$ means ``approximately equal'' and its meaning  is clear from the theorem.

These two parameters $A$, $B$ are   \emph{features of the EEG signal} useful  in the classification of the ``short'' EEG records.  These features don't depend from the data generating mechanism and are model-free.  In the scalar case they have been introduced and applied for the purpose to detect changes in generating mechanism \cite{darkhovsky2012new,darkhovsky2014new,darkhovsky2014quickest,darkhovsky2015novel} of ``long''  EEG data.


\begin{thebibliography}{99}
\bibitem{haynes2006decoding}
J.-D.~Haynes, G.~Rees, Decoding mental states from brain activity in humans,
Nature Reviews Neuroscience 7 (7) (2006) 523-534.


\bibitem{aflalo2015decoding}
 T. Aflalo, S. Kellis, C. Klaes, B. Lee, Y. Shi, K. Pejsa, K. Shanfield,
S. Hayes-Jackson, M. Aisen, C. Heck, et al., Decoding motor imagery from
the posterior parietal cortex of a tetraplegic human, Science 348 (6237)
(2015) 906-910.

\bibitem{kaplan2005unconscious}
 A. Y. Kaplan, J.-G. Byeon, J.-J. Lim, K.-S. Jin, B.-W. Park, S. U. Tarasova,
Unconscious operant conditioning in the paradigm of brain-computer interface
based on color perception, International journal of neuroscience 115 (6)
(2005) 781-802.


\bibitem{Kaplan2013}
 A. Kaplan, S. Shishkin, I. Ganin, I. Basyul, A. Zhigalov, Adapting the
p300-based brain-computer interface for gaming: a review, Computational
Intelligence and AI in Games, IEEE Transactions on 5 (2) (2013) 141-149.

\bibitem{loo2015research}
 S. K. Loo, A. Lenartowicz, S. Makeig, Research review: use of eeg biomarkers
in child psychiatry research--current state and future directions, Journal
of Child Psychology and Psychiatry.

\bibitem{neto2015eeg}
E. Neto, E. A. Allen, H. Aurlien, H. Nordby, T. Eichele, Eeg spectral features
discriminate between alzheimeras and vascular dementia, Frontiers
in neurology 6.

 \bibitem{van2007application}
 O. Van Der Stelt, A. Belger, Application of electroencephalography to the
study of cognitive and brain functions in schizophrenia, Schizophrenia bulletin
33 (4) (2007) 955-970.

\bibitem{boutros2008status}
 N. N. Boutros, C. Arfken, S. Galderisi, J. Warrick, G. Pratt, W. Iacono,
The status of spectral eeg abnormality as a diagnostic test for schizophrenia,
Schizophrenia research 99 (1) (2008) 225-237.

\bibitem{babloyantz1986evidence}
 A. Babloyantz, Evidence of chaotic dynamics of brain activity during the
sleep cycle, in: Dimensions and entropies in chaotic systems, Springer,
1986, pp. 241-245.

\bibitem{roschke1992strange}
 J. R{\"o}oschke, Strange attractors, chaotic behavior and informational aspects
of sleep eeg data, Neuropsychobiology 25 (3) (1992) 172-176.

\bibitem{pereda}
 E. Pereda, A. Gamundi, R. Rial, J. Gonz{\'a}lez, Non-linear behaviour of
human eeg: fractal exponent versus correlation dimension in awake and
sleep stages, Neuroscience letters 250 (2) (1998) 91-94.

\bibitem{kaplan2001}
 A. Kaplan, J. R{\"o}schke, B. Darkhovsky, J. Fell, Macrostructural eeg characterization
based on nonparametric change point segmentation: application
to sleep analysis, Journal of neuroscience methods 106 (1) (2001) 81-90.

\bibitem{subha2010eeg}
 D. P. Subha, P. K. Joseph, R. Acharya, C. M. Lim, Eeg signal analysis: a
survey, Journal of medical systems 34 (2) (2010) 195-212.


\bibitem{Kaplan1998Nonstationarity}
 A. Kaplan, Nonstationarity eeg: methodological and experimental analysis.,
Advances of Physiological Sciences 29 (3) (1998) 35-55.

\bibitem{dvey2015schizophrenia}
Z. Dvey-Aharon, N. Fogelson, A. Peled, N. Intrator, Schizophrenia detection
and classification by advanced analysis of eeg recordings using a single
electrode approach, PloS one 10 (4) (2015) e0123033.

\bibitem{dark2002}
 B. Darkhovskii, A. Kaplan, S. Shishkin, On an approach to the estimation
of the complexity of curves (using as an example an electroencephalogram
of a human being), Automation and Remote Control 63 (3) (2002) 468-474.


\bibitem{darkhovsky2006estimation}
 B. Darkhovsky, A. Kaplan, M. Kosinov, The estimation of complexity for
the electroencephalogram in humans, in: 2006 IEEE Conference on Computer
Aided Control System Design, 2006 IEEE International Conference
on Control Applications, 2006 IEEE International Symposium on Intelligent
Control, 2006.
\bibitem{darkhovsky2006estimation}
 B. Darkhovsky, A. Piryatinska, A new complexity-based algorithmic procedures
for electroencephalogram (eeg) segmentation, in: Signal Processing
in Medicine and Biology Symposium (SPMB), 2012 IEEE, IEEE, 2012, pp.
1-5.
\bibitem{darkhovsky2014new}
 B. Darkhovsky, A. Piryatinska, New approach to the segmentation problem
for time series of arbitrary nature, Proceedings of the Steklov Institute of
Mathematics 287 (1) (2014) 54-67.

\bibitem{darkhovsky2014quickest}
 B. Darkhovsky, A. Piryatinska, Quickest detection of changes in the generating
mechanism of a time series via the "-complexity of continuous functions,
Sequential Analysis 33 (2) (2014) 231-250.

\bibitem{darkhovsky2015novel}
 B. Darkhovsky, A. Piryatinska, Novel methodology of change-points detection
for time series with arbitrary generating mechanisms, in: Stochastic
Models, Statistics and Their Applications, Springer, 2015, pp. 241-251.

\bibitem{breiman2001random}
 L. Breiman, Random forests, Machine learning 45 (1) (2001) 5-32.
\bibitem{geisser1993predictive} 
 S. Geisser, Predictive inference, Vol. 55, CRC Press, 1993.

\bibitem{kohavi1995study}
R. Kohavi, et al., A study of cross-validation and bootstrap for accuracy
estimation and model selection, in: Ijcai, Vol. 14, 1995, pp. 1137-1145.

\bibitem{efron1997improvements}
 B. Efron, R. Tibshirani, Improvements on cross-validation: the 632+ bootstrap
method, Journal of the American Statistical Association 92 (438)
(1997) 548-560.


\bibitem{borisov2004segmental}
 S. Borisov, A. Kaplan, N. Gorbachevskaia, I. Kozlova, Segmental structure
of the eeg alpha activity in adolescents with disorders of schizophrenic
spectrum, Zhurnal vysshei nervnoi deiatelnosti imeni IP Pavlova 55 (3)
(2004) 329-335.


\bibitem{ala2005classification}
 A. Kaplan, S. Borisov, V. Zheligovskii, Classification of the adolescent eeg
by the spectral and segmental characteristics for normals, Zh Vyssh Nerv
Deiat Im IP Pavlova 55 (2005) 478-86.
\bibitem{borisov2005analysis}
 S. Borisov, A. Kaplan, N. Gorbachevskaia, I. Kozlova, Analysis of eeg
structural synchrony in adolescents suffering from schizophrenic disorders,
Fiziologiia cheloveka 31 (3) (2005) 16.

\bibitem{md1994diagnostic}
 W. Maier, Diagnostic and statistical manual of mental disorders.

\bibitem{md1994diagnostic}
 V. R. Badrakalimuthu, R. Swamiraju, H. de Waal, Eeg in psychiatric practice:
to do or not to do?, Advances in psychiatric treatment 17 (2) (2011)
114-121.

\bibitem{kim2015diagnostic}
 J. W. Kim, Y. S. Lee, D. H. Han, K. J. Min, J. Lee, K. Lee, Diagnostic utility
of quantitative eeg in un-medicated schizophrenia, Neuroscience letters
589 (2015) 126-131.


\bibitem{light2015validation}
 G. A. Light, N. R. Swerdlow, M. L. Thomas, M. E. Calkins, M. F. Green,
T. A. Greenwood, R. E. Gur, R. C. Gur, L. C. Lazzeroni, K. H. Nuechterlein,
et al., Validation of mismatch negativity and p3a for use in multi-site
studies of schizophrenia: characterization of demographic, clinical, cognitive,
and functional correlates in cogs-2, Schizophrenia research 163 (1)
(2015) 63-72.



\bibitem{hohne2014motor}
 J. H{\"o}hne, E. Holz, P. Staiger-S{\"a}lzer, K.-R. M{\"u}ller, A. K{\"u}bler, M. Tangermann,
Motor imagery for severely motor-impaired patients: evidence for
brain-computer interfacing as superior control solution.

%\bibliography{mybibl}

\end{thebibliography}
\end{document}